\documentclass[twocolumn,prb,showpacs]{revtex4}

\usepackage{graphicx}
\usepackage{rotating}
\usepackage{amsmath}
\usepackage{amsfonts}
\usepackage{amssymb}
\usepackage{enumerate}
\usepackage{longtable}
\usepackage{dcolumn}
\usepackage{bm}

\usepackage{mathptmx}
\usepackage{bbm}

\begin{document}

\title{Two-Dimensionality of Magnetic Excitations on the 
Trellis Lattice: (La,Sr,Ca)$_{14}$Cu$_{24}$O$_{41}$ and SrCu$_2$O$_3$}

\author{K.P. Schmidt$^1$}
\email{kaiphillip.schmidt@epfl.ch}
\affiliation{$^1$ Institute of Theoretical Physics, 
\'{E}cole Polytechnique F\'{e}d\'{e}rale de Lausanne, CH-1015 Lausanne, 
Switzerland}
\author{G.S. Uhrig$^2$}
\email{goetz.uhrig@uni-dortmund.de}
\affiliation{$^2$ Lehrstuhl f\"ur theoretische Physik I, 
Universit\"at Dortmund, D-44221 Dortmund, Germany}

\date{\rm\today}

\begin{abstract} 
We explore the properties of magnetic excitations on the trellis lattice which 
is relevant for the so-called telephone-number compounds 
A$_{14}$Cu$_{24}$O$_{41}$ and the system SrCu$_2$O$_3$. The trellis lattice 
consists of two-leg ladders which are coupled in a strongly frustrated fashion.
 We use the effective model obtained for a single two-leg spin ladder to 
calculate the two-dimensional one-triplon dispersion and the corresponding 
one-triplon contribution to the dynamical structure factor. Special attention 
is laid on signatures of the frustrating inter-ladder magnetic exchange. A 
detailed suggestion is made for an experimental detection of this exchange in 
inelastic neutron scattering experiments.     
\end{abstract}

\pacs{74.25.Ha, 75.40.Gb, 75.10.Jm, 75.50.Ee}


\maketitle

Low dimensional quantum antiferromagnets display a variety of fascinating 
properties. The reduced dimensionality can either be realized by strongly 
anisotropic magnetic exchanges or by a strongly frustrating topology of the 
system. The so-called telephone-number compounds A$_{14}$Cu$_{24}$O$_{41}$ and 
the system SrCu$_2$O$_3$ are often discussed as realizations of idealized quasi
 one-dimensional two-leg ladders
\cite{dagot96,mizun97,brehm99,windt01,nunne02,schmi05a,notbo07}. These 
one-dimensional structures build a two-dimensional trellis lattice 
(see Fig.~1), i.e. the coupling between the ladders is strongly frustrated. 
But the detailed influence of such a coupling and its quantitative strength 
are currently unknown. 

Quantum chemistry calculations have revealed a sizable value for the magnetic 
exchange between the one-dimensional ladder structures for the system 
SrCu$_2$O$_3$\cite{morei06}. It is therefore an important issue to clarify the 
impact of the two-dimensionality of these systems on their magnetic properties.
Theoretically, this question has been treated by mean-field theory using the 
bond-operator approach which is reliable when the rung-coupling $J_\perp$ 
dominates\cite{gopal94}. Exact diagonalization studies were performed on small 
clusters which provide only a limited number of points in momentum space
\cite{riera01}. In this work we approach this issue by studying the effects of 
the inter-ladder coupling $J_{\rm int}$ on the elementary excitations of the 
single ladder and its dynamics. We concentrate on the paramagnetic case which 
is the experimentally relevant one. We give a detailed description for a 
possible experimental detection and determination of such a coupling by 
inelastic neutron scattering.      

\begin{figure}[htb]
  \centerline{\includegraphics[width=0.8\columnwidth]{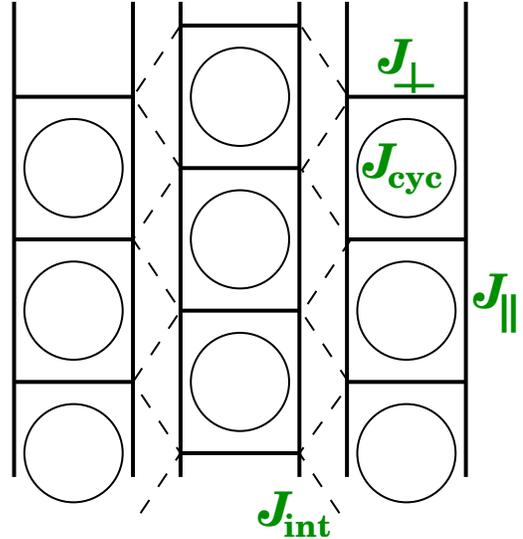}}
  \caption{Heisenberg exchange couplings ($J_\perp, J_\parallel, J_\text{int}
    $) and the four-spin exchange coupling $J_\text{cyc}$ (circles) considered 
in the
    Hamiltonian (\ref{eq:hamilton}). The rung coupling (horizontal solid 
    lines) is $J_\perp$, the one along the legs of the ladders 
    (vertical solid lines) is denoted by $J_\parallel$ and the
    coupling between adjacent ladders (dashed lines) 
    is denoted by $J_\text{int}$.
  }
  \label{fig:coupling}
\end{figure}

\begin{figure}[htb]
  \centerline{\includegraphics[width=0.8\columnwidth]{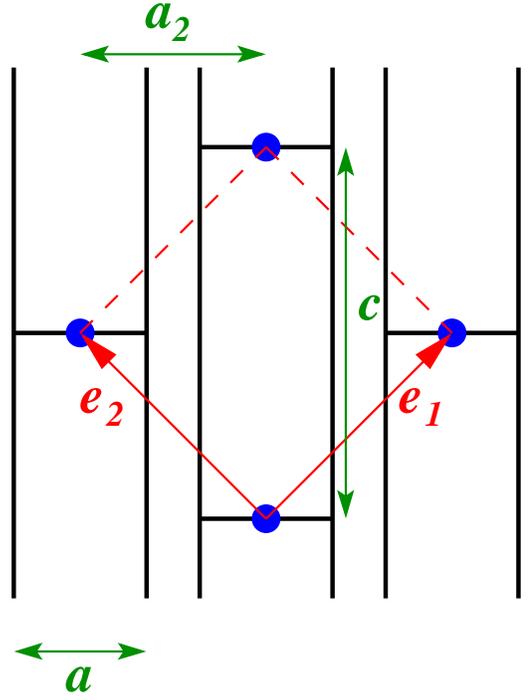}}
  \caption{Unit cell of a trellis lattice spanned by the unit vectors $e_1$
    and $e_2$. The distance between two adjacent 
    rungs on the same ladder is $c$.
    The distance between the legs of a ladder is $a$, the distance
    perpendicular to the ladders between the centers (blue dots) 
    of two rungs on adjacent ladders is $a_2$.}
  \label{fig:unitcell}
\end{figure}

The couplings considered are shown in Fig.\ \ref{fig:coupling}. They refer to 
the Hamiltonian
\begin{subequations}
  \label{eq:hamilton}
  \begin{eqnarray}
    H   &=&  H_0 +H_\text{cyc} +H_\text{int}\\
    H_0 &=& \vspace*{-1mm}J_\perp \sum\limits_{i,j} {\bf S}_{i,j;\text{L}}
            {\bf S}_{i,j;\text{R}}\nonumber\\ 
    &&+J_\parallel\vspace*{-5mm}\sum\limits_{i,j;\tau\in\{\text{L}.\text{R}\}} 
    {\bf S}_{i,j;\tau} {\bf S}_{i+1,j+1;\tau}\\
    H_\text{cyc} &=& J_\text{cyc}\sum_{i,j}\Big[({\bf S}_{i,j;\text{L}}{\bf
S}_{i+1,j+1;\text{L}})({\bf S}_{i,j;\text{R}}{\bf S}_{i+1,j+1;\text{R}}) 
\nonumber\\
      &+& 
      ({\bf S}_{i,j;\text{L}}{\bf S}_{i,j;\text{R}})
      ({\bf S}_{i+1,j+1;\text{L}}{\bf S}_{i+1,j+1;\text{R}})
      \nonumber\\
      && \qquad \quad
      - ({\bf S}_{i,j;\text{L}}{\bf S}_{i+1,j+1;\text{R}})
      ({\bf S}_{i,j;\text{R}}{\bf S}_{i+1,j+1;\text{L}}) \Big] \\
    H_\text{int} &=& J_\text{int}     \sum\limits_{i,j} {\bf S}_{i,j;\text{R}}
    \Big[ {\bf S}_{i+1,j;\text{L}}+{\bf S}_{i,j-1;\text{L}}\Big]
  \end{eqnarray}
\end{subequations}
where the pair $i, j \in \mathbbm{Z}$ denotes the rungs, not the sites, 
counted along the unit vectors $e_1$ and $e_2$ shown in 
Fig.~\ref{fig:unitcell}. The 
subscripts L and R stand for the left and right spin, respectively, on the 
particular rung. 

In order to explain the influence of various changes of parameters we give 
below the formulae for the two-dimensional dispersion $\omega_{k,q}$ on a 
trellis lattice and for the dynamic structure factor $S_{k,q}(\omega)$. The 
approach starts from the effective model that we have derived previously for a 
single ladder\cite{knett01b,schmi05b}. It is given in terms of triplons, the 
elementary excitations of the spin ladder. 
In the following, we give a summarized description how we derived the 
effective model for the single spin ladder. 
Technical details are given in Refs.\ \onlinecite{knett03a,knett04b}. 

We apply a particle-conserving perturbative continuous unitary transformation 
(CUT) which uses the states on isolated rungs as reference. The CUT maps the 
Hamiltonian of the single ladder to an effective Hamiltonian which conserves 
the number of triplons. In order to calculate spectral properties and to treat 
the inter-ladder coupling $H_{\rm int}$ the relevant observables have to be 
transformed by the same CUT. Here we use the perturbative realization of CUT 
which gives series expansions for the effective Hamiltonian and the effective 
observables in terms of the expansion parameters $x:=J_\parallel/J_\perp$ and 
$x_{\rm cyc}:=J_{\rm cyc}/J_\perp$ in the thermodynamic limit. The obtained 
series are extrapolated giving reliable results in a broad range of the 
parameter space\cite{schmi03a,schmi03d}.         

For the present purpose, we need to know $\omega^0_k$, the single-triplon 
dispersion along the ladder and the spectral weight $a_k^2$ of the single 
triplon when it is excited by $S^z$ or another spin component. 
This spectral weight 
is known as function of the wave vector $k$. The coupling between the ladders 
is treated on the mean-field level starting from the ladders, not the isolated 
rung dimers\cite{gopal94}, as unperturbed system\cite{uhrig04a,uhrig05a}. We 
stress that the hardcore property of the triplons is taken into account along 
the ladders but not perpendicular to them. 

The rungs of the trellis lattice form a Bravais lattice so that there is 
exactly one mode of given magnetic quantum number $S^z$ per point in the first 
Brillouin zone. But the lattice has a skewed unit cell due to the fact that 
adjacent ladders are shifted with respect to each other by half the distance 
$c$ between two rungs in one ladder (see Fig.~2). Conventionally, one would 
express the momentum in units of the reciprocal basis vectors $e_1^*$ and 
$e_2^*$. But in this system it is more convenient to describe the momentum by 
its component along the ladders ($k$, measured in units of $2\pi/c$) and by its
 component perpendicular to the ladder ($q$, measured in units of $2\pi/(2a_2)$
 , cf.\ Fig.\ \ref{fig:unitcell}). 
Note that as a consequence the dispersion 
$\omega(k,q)$ does not have the periodicities one would naively expect. 
The periodicity in $k$ and $q$ separately is 2, not unity.
But shifting $k\to k+1$ and $q\to q+1$ together reproduces the spectrum
as it has to be.\footnote{The skewed unit cell as shown in Fig.\ 
\ref{fig:unitcell} is simpler in spite of the `unusual' periodicities
because it contains only one dimer so that it is a Bravais lattice with one
triplon mode per unit cell. If one
insisted on using unit vectors along and perpendicular to the ladders the unit
cell would comprise two dimers so that two modes per unit cell have to
be considered. They correspond to the modes at $q$ and at $q+1$.}

We use the same approach as in Refs.\ \onlinecite{uhrig04a,uhrig05a} to treat 
the coupling $H_{\rm int}$ between the ladders. The observables 
$S_{i}^{x,y,z,\text{R}}$ are transformed according to
\begin{equation}
\label{obs-trafo1}
S_{i,\text{eff}}^{x,y,z,\text{R}}
:= U^\dagger S_{i}^{x,y,z,\text{R}} U
=
\sum_\delta a_\delta \, (t^{x,y,z,\dagger}_{i+\delta} + t^{x,y,z}_{i+\delta})
+ \ldots
\end{equation}
where the dots stand for normal-ordered quadratic and higher terms in the 
bosonic operators. In the following we neglect these terms corresponding to 
multi-triplon contributions. Note that the symmetry 
$S_{i,\text{eff}}^{\alpha,\text{L}} \!  = 
\!-S_{i,\text{eff}}^{\alpha,\text{R}}$ holds on this level of approximation. 
The Fourier transform squared of $a_\delta$ yields the one-triplon spectral 
weight $a_k^2$. In this approximation, the total Hamiltonian of the trellis 
lattice is quadratic in terms of the operators $t^{x,y,z,\dagger }$ and 
$t^{x,y,z}$. Neglecting the hardcore property allows us to diagonalize the 
Hamiltonian by a Bogoliubov transformation which is justified as long as the 
coupling $J_{\rm int}$ is small.     

The following two-dimensional dispersion $\omega_{k,q}$ is obtained
\begin{equation}
\omega_{k,q} = \omega^0_{k} \sqrt{1-\frac{8J_\text{int}}{\omega_k^0}
a^2_k \cos(\pi q )\cos(\pi k)}
\ .
\label{eq:dispersion}
\end{equation}
The two-dimensional dynamic structure factor reads
\begin{subequations}
\begin{eqnarray}
S_{k,q}(\omega) &=&  a^2_{k,q}\, \delta(\omega-\omega_{k,q})\\
a^2_{k,q} &=& 2\sin^2\left(\pi q\frac{a}{2a_2}\right) a^2_k \, 
\frac{\omega_k^0}{\omega_{k,q}}
\ .
\end{eqnarray}
\label{eq:weight}
\end{subequations}
The sine factor stems from the interference of the excitation processes from 
the left and from the right spin in each rung. Recall that a single triplon has
 an odd parity on the isolated ladder with respect to reflections about the 
centerline\cite{knett01b,schmi05b}.

\begin{figure}[htb]
  \centerline{\includegraphics[width=\columnwidth]{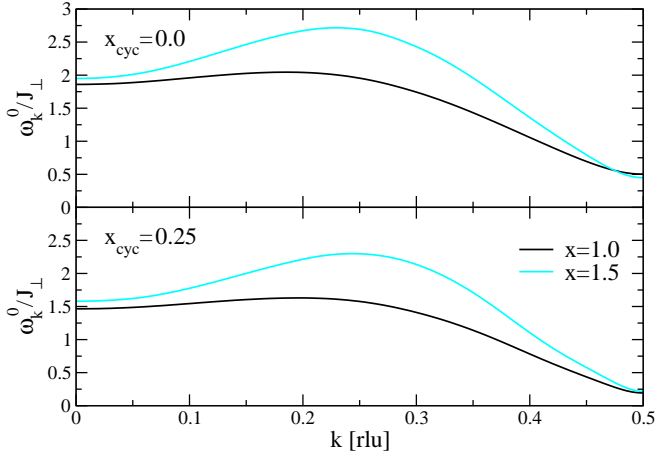}}
  \caption{Dispersion $\omega^0_k$ of a single triplon for an isolated ladder. 
{\it Upper panel:} One-triplon dispersion for $x_{\rm cyc}=0$ and $x=1$ (black 
curve) and $x=1.5$ (cyan/grey curve). {\it Lower panel:} One-triplon dispersion
 for $x_{\rm cyc}=0.25$ and $x=1$ (black curve) and $x=1.5$ (cyan/grey curve).}
  \label{fig:dispersion}
\end{figure}

\begin{figure}[htb]
  \centerline{\includegraphics[width=\columnwidth]{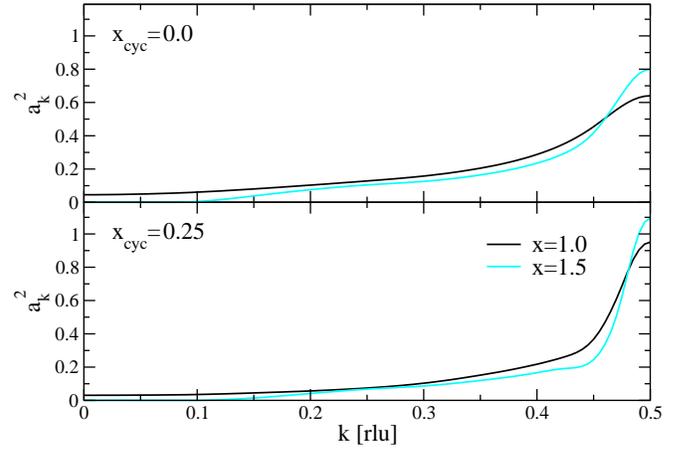}}
  \caption{Weight $a^2_k$ of a single triplon for an isolated ladder. 
{\it Upper panel:} One-triplon spectral weight for $x_{\rm cyc}=0$ and $x=1$ 
(black curve) and $x=1.5$ (cyan/grey curve). {\it Lower panel:} One-triplon 
spectral weight for $x_{\rm cyc}=0.25$ and $x=1$ (black curve) and $x=1.5$ 
(cyan/grey curve).}  \label{fig:weight}
\end{figure}

The data for an isolated ladder is shown in Fig.~\ref{fig:dispersion} for the 
one-triplon dispersion of a single ladder and in Fig.~\ref{fig:weight} for the 
one-triplon weight. The couplings and energies are given in units of the rung 
coupling $J_\perp$, that is $x:=J_\parallel/J_\perp$, $x_\text{cyc}:=
J_\text{cyc}/J_\perp$. This constitutes the input data for the two-dimensional 
calculation. 


Note that at $x=1.5$ the extrapolation of the weight at small momentum 
spuriously leads 
 to slightly negative values. In this parameter regime the 
three-triplon continuum overlaps with the one-triplon dispersion leading to a 
possible decay  and a finite life-time of the elementary excitations. 
Therefore we put the one-triplon spectral weight in this range to zero.
\footnote{Recall that the overall total spectral weight at small momentum is 
small anyway. Additionally, the three-triplon contribution to the dynamical 
structure 
factor behaves like  $\omega^{3}$ close to the lower band edge leading to a 
long life-time and therefore a sharp resonance at small momentum.
This high power results from the hardcore property of the triplons.} 
We refrain from displaying data for even larger values of $x$ because
there is a growing body of evidence that $x\in (1,1.5)$ is the relevant
range for cuprate ladders 
\cite{matsu00a,matsu00b,windt01,nunne02,schmi05a,notbo07}
even though early quantum chemical calculations \cite{johns00a}
indicated even larger values of
 $x\approx2$. Moreover, our approach is less reliable beyond  $x\approx 1.5$.

\begin{figure}[htb]
  \centerline{\includegraphics[width=\columnwidth]{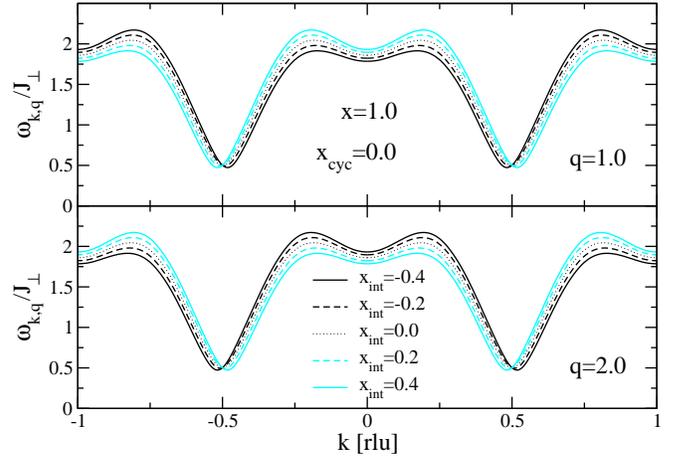}}
  \caption{The influence of the inter-ladder coupling $x_{\rm int}$ on the 
dispersion $\omega_{k,q}$ of a single triplon on the trellis lattice as shown 
in Figs.~\ref{fig:unitcell} for $x=1$ and $x_{\rm cyc}=0$. {\it Upper panel:} 
Data sets for $q=1$ and $x_{\rm int}=\{-0.4;-0.2;0.0;0.2;0.4\}$. {\it Lower 
panel:} Data sets for $q=2$ and $x_{\rm int}=\{-0.4;-0.2;0.0;0.2;0.4\}$.}
  \label{fig:disp_trellis_x1.0_div_xint}
\end{figure}

In the following, we will first discuss the generic properties of the 
two-dimensional spectra. To this end, we concentrate on the case $x=1$ and 
$x_{\rm cyc}=0$. But the features obtained are similar for all couplings in the
 paramagnetic phase. The influence of the inter-ladder coupling $x_{\rm int}$ 
on the dispersion of a single triplon is shown in 
Fig.~\ref{fig:disp_trellis_x1.0_div_xint}. For $q=1.5$ the effect of the 
trellis structure vanishes completely due to destructive interference.
The largest effects can be seen for $q=1$ and $q=2$. 
(We refer to $q=1.5$ and $q=2$ instead of $q=0.5$ and $q=0$ although this does 
not make any difference in the framework of our theoretical model. But it eases
the contact to the experimental investigations, cf.\ e.g.\ 
Ref.~\onlinecite{notbo07}, where one keeps away from zero momentum where no 
magnetic scattering occurs.)
 Clearly, deviations from $q=1.5$ lead to a lifting 
of the reflection symmetry about $k=\pm 0.5$ and to a difference
between $q=1$ and $q=2$. On the level of our description, a ferromagnetic 
coupling $-|x_{\rm int}|$ at  $q=1$ is equivalent to an antiferromagnetic 
coupling $|x_{\rm int}|$ at $q=2$ 
and vice versa. The one-triplon gap is shifted for an antiferromagnetic 
coupling from $k=0.5$ to a \emph{lower} momentum for $q=2$ and from $k=0.5$ to 
a  \emph{higher} momentum for $q=1$. 

\begin{figure}[htb]
  \centerline{\includegraphics[width=\columnwidth]{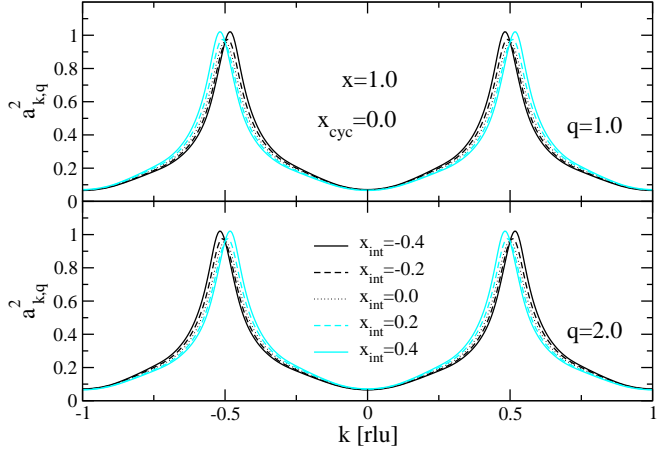}}
  \caption{The influence of the inter-ladder coupling $x_{\rm int}$ on the 
spectral weight $a_{k,q}^2$ of a single triplon on the trellis lattice as shown
 in 
Figs.~\ref{fig:unitcell} for $x=1$ and $x_{\rm cyc}=0$. {\it Upper panel:} Data
 sets for $q=1$ and $x_{\rm int}=\{-0.4;-0.2;0.0;0.2;0.4\}$. {\it Lower panel:}
 Data sets for $q=2$ and $x_{\rm int}=\{-0.4;-0.2;0.0;0.2;0.4\}$.}
  \label{fig:weight_trellis_x1.0_div_xint}
\end{figure}

The physical interpretation of the destructive interference at half-integer
values of $q$ or  of $k$ relies on the geometry of the trellis lattice, 
see Fig.\ \ref{fig:unitcell}. Consider, for instance, two hopping processes 
from the two rungs, where the arrows $e_1$ and $e_2$ end, to the rung, from 
where both arrows start. For half-integer value of $q$ these processes have 
opposite sign so that they cancel each other. Hence no effect of 
$J_\text{int}$ occurs.
The same is true for two hopping processes from the two
rungs in the ladder in the middle in Fig.\ \ref{fig:unitcell}
to one of the other rungs. For half-integer value of $k$ these processes have 
opposite sign so that they cancel each other. 
Again, no effect of $J_\text{int}$ occurs.
This explains why $J_\text{int}$ has to be significantly large in order
to imply a sizeable effect if at least one of the momenta  $q$ and $k$
is close to a half-integer values. Additionally, Fig.\ \ref{fig:weight}
shows that for values of $k$ which deviate substantially from half-integer
values the spectral weight of the triplons is very low so that the inter-ladder
coupling has only very limited influence on their dynamics,
even away from half-integer values of momentum $k$.

Besides the energy of the elementary triplon excitation, also the spectral 
weight is affected by the inter-ladder coupling. The correspoding results are 
shown in Fig.~\ref{fig:weight_trellis_x1.0_div_xint}. 
In accordance to the one-triplon gap, the maximum of the one-triplon spectral 
weight is shifted 
for an antiferromagnetic coupling from $k=0.5$ to a lower momentum $k$ for 
$q=2$  and  from $k=0.5$ to a higher momentum $k$ for $q=1$.  But the overall 
effect on this quantity is rather small due to the frustration.  

Next we discuss a possible experimental detection of the inter-ladder coupling 
in inelastic neutron scattering experiments. From the above, it becomes clear 
that measurements collecting events in the interval  $q\in[1,2]$ do not allow 
to establish the inter-ladder coupling. We consider the possible detection of 
an inter-ladder coupling to be an essential question because it concerns the
dimensionality of the magnetic systems
A$_{14}$Cu$_{24}$O$_{41}$ and SrCu$_2$O$_3$. Therefore, we suggest to 
measure around $q=1$ and around $q=2$. The comparison of the two 
constant energy scans of $k$  provide the data to determine the 
value of $J_\text{int}$. We point out that the weight factor 
$\sin^2\left(\pi q\frac{a}{2a_2}\right)$ is reduced only by 25\% if one 
deviates from the optimum value of $q=1.5$ (where
the sine is unity) to $q=1$ or $q=2$. 
In this argument, we assume the approximate ratio $a_2=1.5a$.

\begin{figure}[htb]
  \centerline{\includegraphics[width=\columnwidth]{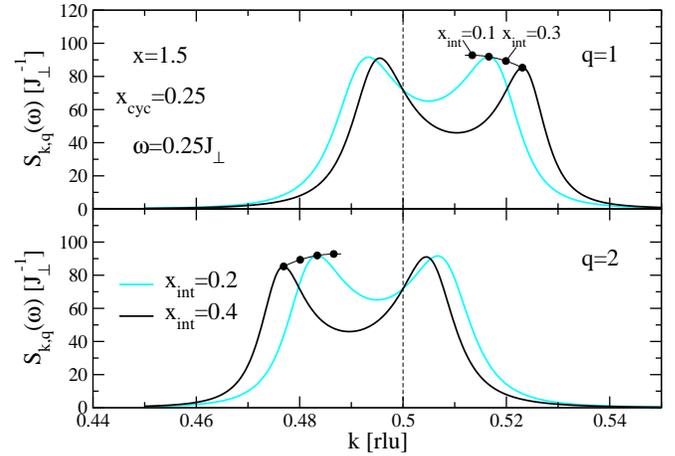}}
  \caption{Constant $\omega$-scan of the dynamical structure factor at 
$\omega=0.25J_\perp$ 
with $x=1.5$ and $x_{\rm cyc}=0.25$ as a function of the momentum
 $k$ in ladder direction. The black curves correspond to $x_{\rm int}=0.4$ and 
the cyan/grey curves correspond to $x_{\rm int}=0.2$. The uppper panel shows 
$q=1$ and the lower panel shows $q=2$. A Lorentzian broadening of $\eta=0.05$ 
is used in  all panels.}
  \label{fig:lorenz_trellis}
\end{figure}

In our opinion, the best experimental way to determine the inter-ladder 
coupling  by inelastic neutron scattering is to perform constant energy scans 
$S(k,q=\pm 1,\omega=\Delta+\delta)$ at momenta $q= 1$  and $q= 2$ for energies 
$\Delta+\delta$ slightly above the one-triplon gap $\Delta$, i.e.\ $\delta>0$. 
The resulting spectra will show characteristic asymmetries with respect to 
reflections  about $k=0.5$, see Fig.~\ref{fig:lorenz_trellis}. 

The existing experimental data of magnetic excitations in 
A$_{14}$Cu$_{24}$O$_{41}$ and SrCu$_2$O$_3$ are consistent with coupling 
constants $x=1.2-1.5$ and 
$x_{\rm cyc}=0.2-0.25$\cite{nunne02,schmi05a,notbo07}. In the following, we 
discuss the parameter set $x=1.5$ and $x_{\rm cyc}=0.25$ 
which has been deduced from recent inelastic neutron scattering data for 
La$_4$Sr$_{10}$Cu$_{24}$O$_{41}$\cite{notbo07}. Typical curves are shown in 
Fig.~\ref{fig:lorenz_trellis}. Note again that in our mean-field treatment 
a change from antiferromagnetic 
to ferromagnetic inter-ladder coupling $|x_{\rm int}| \to - |x_{\rm int}|$ 
leads to the same  curves as in Fig.~\ref{fig:lorenz_trellis} where 
the momentum $q$ is changed to $q+1$.   

The large anisotropy $x=J_\parallel/J_\perp = 1.5$ is questioned by 
quantum chemistry calculations\cite{morei06}. It is argued that the magnetic 
exchanges along the 
legs $J_{\parallel}$ and the rungs $J_{\perp}$ are almost equal but a strong 
ferromagnetic inter-ladder coupling $x_{\rm int}\approx -0.2$ is present (the 
calculations were done for the compound SrCu$_2$O$_3$). We want to point out 
that this scenario of an isotropic exchange $x\approx 1$ is {\it not} 
consistent with our results. As discussed above, the effects of the 
inter-ladder coupling are washed out if one averages over the momentum $q$ 
perpendicular to the ladder direction. This has been done when fitting the 
experimental data in Ref.~\onlinecite{notbo07}. Therefore, the deduced 
parameters $x=1.5$  and $x_{\rm cyc}=0.25$ will not change if one takes into 
account a  finite coupling $x_{\rm int}$. 

Independently from the intra-ladder anisotropy $x$, the strong ferromagnetic 
inter-ladder coupling deduced in Ref.~\onlinecite{morei06} is an interesting 
new aspect of the physics in these materials. It will be interesting to 
see if the analysis of inelastic neutron scattering data using the results 
obtained in this work will yield values of the magnetic exchange 
$J_{\rm int}$ which are consistent with the quantum chemistry calculations.  

In summary, we have discussed the properties of magnetic excitations on the 
frustrated trellis lattice in the paramagnetic phase by calculating the 
one-triplon  dispersion and the corresponding one-triplon contribution to the 
dynamical  structure factor. The trellis lattice can be viewed as being built
from coupling components of one-dimensional two-leg ladders. 
Technically, an effective model for a single two-leg ladder is used as the 
starting point to discuss the influence of the inter-ladder coupling. The 
strong frustration of the lattice causes the system to be effectively 
one-dimensional, i.e.\ the magnetic excitation spectrum reveals many 
properties of a single two-leg spin ladder. The leading order of the
inter-ladder coupling $J_\text{int}$ interferes destructively around
the minima of the dispersion in the ladder. Hence, the effects of small or 
even sizable  inter-ladder couplings are generically small.
Yet, they lead to typical 
asymmetries of the dispersion and the spectral weight of a single mode. 

Although the mobility of triplons is strongly reduced by frustration, we 
expect that two- or multi-triplon properties like the \emph{interaction} 
between the triplons are affected in a stronger fashion. While the
hopping of triplons cancels due to frustration their interaction as
induced by different couplings adds. Thus, it is  an important challenge for
future work to clarify the influence of the inter-ladder coupling on optical 
properties and on all other probes of multi-triplon properties.  

We have made a detailed prediction how to  detect the magnetic inter-ladder 
coupling in experimental systems like the so-called telephone-number compounds 
A$_{14}$Cu$_{24}$O$_{41}$ and the system SrCu$_2$O$_3$. We have provided
arguments that experimental averages over the momentum perpendicular to the 
ladder direction do not yield  information about the inter-ladder coupling. 
Therefore, the recently deduced parameter set $x=1.5$ and $x_{\rm cyc}=0.25$
will not be affected  by a finite inter-ladder coupling \cite{notbo07}.
This experimentally established difference between $J_\parallel$ and
$J_\perp$ calls for further quantum chemical analyses of the superexchange
couplings.

The study of the 
inelastic neutron scattering spectra at $q=1$ and $q=2$ should clarify the 
sign and the size of the two-dimensional inter-ladder coupling in the 
experimental systems. 
Thereby, the imporant issue of the dimensionality of these 
magnetic  systems will be elucidated.

We thank B.~B\"uchner, A.~G\"ossling, B.~Lake, A.~L\"auchli, S.~Notbohm, and 
A.~Tennant for  stimulating and helpful discussions.


\end{document}